\documentclass[
twocolumn,
amsmath,
amssymb,
bm,
aps,
]{revtex4-2}

\usepackage{graphicx}
\usepackage{dcolumn}
\usepackage{bm}

\usepackage[utf8]{inputenc}
\usepackage[T1]{fontenc}
\usepackage{mathptmx}
\usepackage{xcolor}

\usepackage{comment}

\usepackage{ulem}
\newcommand{\COM}[1]{{{\color{magenta}[#1] }}}
\renewcommand{\COM}[1]{}

\begin{document}

\title{Phase autoencoder for rapid data-driven synchronization of rhythmic spatiotemporal patterns}
\author{Koichiro Yawata}
\thanks{Corresponding author: koichiro.yawata.rt@gmail.com}
\author{Ryo Sakuma}
\affiliation{Department of Systems and Control Engineering, Institute of Science Tokyo, Tokyo 152-8552, Japan}
\author{Kai Fukami}
\affiliation{Department of Aerospace Engineering, Tohoku University, Sendai 980-8579, Japan}
\author{Kunihiko Taira}
\affiliation{Department of Mechanical and Aerospace Engineering, University of California, Los Angeles,
CA 90095, USA}
\author{Hiroya Nakao}
\affiliation{Department of Systems and Control Engineering and Research Center for Autonomous Systems Materialogy, Institute of Science Tokyo, Tokyo 152-8552, Japan}

\date{\today}

\begin{abstract}
We present a machine-learning method for data-driven synchronization of rhythmic spatiotemporal patterns in reaction-diffusion systems.
Based on the phase autoencoder [Yawata {\it et al.}, Chaos {\bf 34}, 063111 (2024)], we map high-dimensional field variables of the reaction-diffusion system to low-dimensional latent variables characterizing the asymptotic phase and amplitudes of the field variables.
This yields a reduced phase description of the limit cycle underlying the rhythmic spatiotemporal dynamics in a data-driven manner.
We propose a method to drive the system along the tangential direction of the limit cycle, enabling phase control without inducing amplitude deviations.
With examples of 1D oscillating spots and 2D spiral waves in the FitzHugh-Nagumo reaction-diffusion system,
we show that the method achieves rapid synchronization in both reference-based and coupling-based settings.
These results demonstrate the potential of data-driven phase description based on the phase autoencoder
 for synchronization of  high-dimensional spatiotemporal dynamics.
\end{abstract}

\maketitle

\section{Introduction}

Spatially extended populations of autonomous dynamical elements are widely present in the real world, such as chemical reactions and biological cells. They exhibit a variety of spatiotemporal pattern dynamics that are often relevant to the functioning of individual systems~\cite{Winfree2001geometry}.
In particular, target and spiral waves are typical collective dynamics in excitatory or oscillatory reaction-diffusion (RD) systems~\cite{Winfree2001geometry,mikhailov2006control,zykov2005wave}, which are experimentally observed, for example, in the Belousov-Zhabotinsky reaction and in cardiac tissue.
Such stable rhythmic spatiotemporal patterns can be interpreted as stable limit-cycle oscillations in high-dimensional dynamical systems.

In analyzing limit-cycle oscillators, phase reduction and, more recently, phase-amplitude reduction, have been considered~\cite{Winfree2001geometry,Kuramoto1984chemical,Hoppensteadt1997weakly,Brown2004phase,Ermentrout2010mathematical,wedgwood2013phase,Nakao2016phase,wilson2016isostable,shirasaka2017phase,monga2019phase,Kuramoto2019,Ermentrout2019,kotani2020nonlinear,takata2021fast,namura2024optimal,fujii2025optimal}.
These methods systematically approximate the dynamics of multidimensional limit-cycle oscillators using only the phase along the limit cycle, or additionally using the amplitudes characterizing deviations from the limit cycle.
The simplicity of the reduced phase or phase-amplitude equations facilitates the analysis and control of limit-cycle oscillators, enabling applications such as optimal injection locking (entrainment)~\cite{moehlis2006optimal,zlotnik2013optimal,Monga2019optimal,takata2021fast,kato2021optimization,wilson2022recent,godavarthi2023optimal,mircheski2023phase,namura2024optimal,fujii2025optimal}.
They have also been generalized to limit-cycle oscillations in spatially extended systems~\cite{kawamura2013collective,nakao2014phase,kawamura2015phase,taira2018phase,nair2021phase,nakao2021phase,loe2021phase,kawamura2022adjoint}.

For phase reduction, the asymptotic phase function, or its gradient evaluated on the limit cycle, is required (and also the amplitude function or its gradient for phase-amplitude reduction).
These quantities can be obtained if the mathematical model of the oscillator is known.
However, they need to be estimated from data when the mathematical model is not available. 
Various methods have been proposed to evaluate the asymptotic phase, such as those based on the Hilbert transform~\cite{Pikovsky2001synchronization,kralemann2013vivo,stankovski2015coupling,stankovski2017coupling,arai2022extracting,matsuki2023extended,rosenblum2023inferring,furukawa2024bayesian}, Poincar\'e sections~\cite{arai2025setting}, polynomial regression~\cite{namura2022estimating}, and Gaussian processes~\cite{yamamoto2025gaussian}.
Extended Dynamic Mode Decomposition~\cite{williams2015data,kutz2016dynamic,li2017extended,klus2020data,takata2023definition}, which is a data-driven method for estimating Koopman eigenfunctions, has also been used for this purpose, as the asymptotic phase and amplitudes are closely related to Koopman eigenfunctions~\cite{mauroy2013isostables,mauroy2016global,shirasaka2017phase,mauroy2018global,Kuramoto2019,nakao2020spectral,takata2021fast}.

Recently, data-driven machine learning methods for analyzing dynamical systems and fluid flows have been widely studied~\cite{lusch2018deep,Inubushi2020,fukami2023grasping,bakarji2023discovering,fukagata2023reduced,fukami2024data,yawata2024phase,fukagata2025compressing,hiruta2025,Takeishi2017,Yeung2019,Champion2019,Azencot2020,Li2019,Berman2023,Han2021,naiman2023generative,fu2021data,brunton2020machine,vinuesa2023transformative}.
In~\cite{fukami2024data}, an autoencoder-based approach for phase-amplitude reduction of fluid flow was developed, which maps the high-dimensional oscillating flow field to limit-cycle dynamics in a low-dimensional latent space.
A set of phase-amplitude equations was then obtained by sparse regression of the latent-space dynamics and used to control vortex gust-airfoil interactions.
Also, in~\cite{yawata2024phase}, a `phase autoencoder' was proposed, which uses a physics-informed autoencoder to map the system state of a limit-cycle oscillator to latent variables directly representing the asymptotic phase and amplitudes, yielding a reduced phase description from time-series data. 
Noteworthy here is that the phase autoencoder can decode (i.e., reconstruct) the spatiotemporal pattern of the original system from the latent phase variable, not only encoding the system dynamics in the latent space.

In this study, we extend the phase autoencoder framework to extract low-dimensional phase dynamics from high-dimensional field variables of rhythmic spatiotemporal patterns. 
Moreover, we propose a `tangential driving' method for phase control of spatiotemporal patterns, which uses the phase autoencoder to drive the system state tangentially along the limit cycle, thereby suppressing amplitude deviations.
This method does not require accurate estimation of the gradients of the phase and amplitude functions from data, which becomes particularly challenging for high-dimensional spatiotemporal dynamics.
Using the 1D oscillating spots and 2D spiral waves in the FitzHugh-Nagumo RD systems as examples, we demonstrate that our data-driven method rapidly synchronizes spatiotemporal patterns in both reference-based and coupling-based settings.

This paper is organized as follows. In Sec.~2, we cover the phase-amplitude description for RD systems exhibiting limit-cycle oscillations.
In Sec.~3, we present the architecture of the phase autoencoder for spatiotemporal patterns and tangential driving method for data-driven synchronization.
In Sec.~4, we apply the method to two typical patterns of the FitzHugh-Nagumo RD system to demonstrate its effectiveness for synchronizing such patters.
Section~5 concludes the paper.

\section{Phase-amplitude description}

Here, we briefly review the phase-amplitude description for limit-cycle oscillators~\cite{Winfree2001geometry,Kuramoto1984chemical,Hoppensteadt1997weakly,Brown2004phase,Ermentrout2010mathematical,Nakao2016phase,monga2019phase,Kuramoto2019,Ermentrout2019,wilson2016isostable,shirasaka2017phase,takata2021fast} and its extension to RD systems exhibiting rhythmic spatiotemporal patterns~\cite{nakao2014phase,kawamura2017optimizing}.
For simplicity, we first consider a limit-cycle oscillator described by an ordinary differential equation,
\begin{align}
\dot{\bm X}(t) = {\bm F}({\bm X}(t)),
\end{align}
where ${\bm X}(t) \in {\mathbb R}^N$ is the system state at time $t$, $(\dot{})$ represents time derivative, and ${\bm F} : {\mathbb R}^N \to {\mathbb R}^N$ is a smooth vector field representing the system dynamics.
We assume that ${\bm F}$ has an exponentially stable limit-cycle solution ${\bm X}_0(t) = {\bm X}_0(t+T)$ with a natural period $T$.
We can then introduce an asymptotic phase function $\Theta({\bm X}) : B \to [0, 2\pi)$
and $N-1$ amplitude functions $R_j({\bm X}) : B \to {\mathbb C}$ ($j=2, ..., N$)
in the basin $B \subseteq {\mathbb R}^N$ of the limit cycle, satisfying
\begin{align}
\label{eq:phaseampode}
&\dot{\Theta}({\bm X}) = \nabla \Theta({\bm X}) \cdot \dot{\bm X} =  \nabla \Theta({\bm X})  \cdot {\bm F}({\bm X}) =  \omega, \cr
\quad
&\dot{R}_j({\bm X}) = \nabla R_j({\bm X}) \cdot \dot{\bm X} = \nabla R_j({\bm X}) \cdot  {\bm F}({\bm X}) = \lambda_j R_j({\bm X}), 
\quad\quad
\end{align}
where $\nabla$ represents the gradient operator, $(\cdot)$ is the dot product of two vectors~\cite{Kuramoto1984chemical,Nakao2016phase}, 
$\omega = 2\pi / T$ is a natural frequency of the limit cycle, and $\lambda_j$ is the $(j+1)$th Floquet exponent of the limit cycle with a negative real part. The Floquet exponents are sorted in decreasing order of the real parts as $0 > \mbox{Re}\ \lambda_1 \geq \mbox{Re}\ \lambda_2 \geq \cdots$.
The asymptotic phase constantly increases with the frequency $\omega$, and each amplitude decreases exponentially with the decay rate $\mbox{Re}\ \lambda_j$.
In what follows, we define the phase of the system as $\theta = \Theta({\bm X})$, and represent the system state on the limit cycle as ${\bm \chi}(\theta) = {\bm X}_0(\theta / \omega)$ as a function of $\theta$. The amplitude vanishes on the limit cycle, i.e., $R_j({\bm \chi}(\theta)) = 0$ for any $\theta$.

We assume that the system state is close to the limit cycle at time $t$, i.e., ${\bm X}(t) = {\bm \chi}(\theta(t)) + O(\varepsilon)$ with phase $\theta(t) = \Theta({\bm X}(t))$, where $0 \leq \varepsilon \ll 1$ is a small parameter. Then, each amplitude is small, i.e., $r_j(t) = R_j({\bm X}(t)) = R_j( {\bm \chi}(\theta(t)) + O(\varepsilon) ) = O(\varepsilon)$ since $R_j({\bm \chi}(\theta)) = 0$. 
When this oscillator is driven by an additional driving input ${\bm p}(t) \in {\mathbb R}^N$
as 
\begin{align}
\dot{\bm X}(t) = {\bm F}({\bm X}(t)) + {\bm p}(t),
\end{align}
the phase and amplitudes obey 
\begin{align}
\dot{\theta}(t) &= \omega + {\bm Z}({\theta}(t)) \cdot {\bm p}(t), \cr
\dot{r_j}(t) &= \lambda_j r_j(t)+ {\bm I}_j({\theta}(t)) \cdot {\bm p}(t), \quad (j=2, ..., N),
\end{align}
up to first order in $\varepsilon$, where ${\bm Z}(\theta) = \nabla \Theta({\bm \chi}(\theta))$
is the phase sensitivity function and ${\bm I}_j(\theta) = \nabla R_j({\bm \chi}(\theta))$
is the $j$th amplitude sensitivity function (a.k.a. isostable response functions), 
given by the gradients of $\Theta$ and $R_j$ evaluated
at the system state ${\bm \chi}_0(\theta)$ with phase $\theta$ on the limit cycle~\cite{wilson2016isostable,shirasaka2017phase,monga2019phase,takata2021fast}.
Here, ${\bm p}(t)$ need not necessarily be weak as long as the system state stays in the $O(\varepsilon)$ vicinity of the limit cycle as assumed above
(see~\cite{takata2021fast} for details).

In this study, rather than considering a general input ${\bm p}(t)$, we consider {\it tangential driving} and assume that the oscillator is driven only in the tangential direction ${\bm \chi}'(\theta) = d{\bm \chi}(\theta) / d\theta$ of the limit cycle as
\begin{align}
\dot{\bm X}(t) = {\bm F}({\bm X}(t)) + p(t) {\bm \chi}'(\theta(t)),
\end{align}
where $p(t) \in {\mathbb R}$ is the driving amplitude. 
Then, the phase obeys
\begin{align}
\dot{\theta}(t) 
&= \nabla \Theta({\bm X}(t))  \cdot {\bm F}({\bm X}(t)) + p(t) \nabla \Theta({\bm X}(t)) \cdot {\bm \chi}'(\theta(t)) \cr
&= \omega + p(t)  {\bm Z}(\theta(t)) \cdot {\bm \chi}'(\theta(t)) + O(\varepsilon)  \cr
&= \omega + p(t) + O(\varepsilon),
\end{align}
because 
${\bm Z}(\theta) \cdot {\bm \chi}'(\theta) =  \nabla \Theta({\bm X}_0(t')) \cdot \omega^{-1} \dot{\bm X}_0(t') = \omega^{-1} \nabla \Theta({\bm X}_0(t')) \cdot {\bm F}({\bm X}_0(t'))= 1$ from Eq.~(\ref{eq:phaseampode}), where $t' = \theta / \omega$.
Similarly, the amplitude obeys
\begin{align}
\dot{r}_j(t) 
&= \nabla R_j({\bm X}(t)) \cdot {\bm F}({\bm X}(t)) + p(t) \nabla R_j({\bm X}(t)) \cdot {\bm \chi}'(\theta(t)) \cr
&= \lambda_j R_j({\bm X}(t)) + p(t) {\bm I}_j(\theta(t)) \cdot {\bm \chi}'(\theta(t)) + O(\varepsilon) \cr
&= \lambda_j r_j(t) + O(\varepsilon)
\end{align}
since 
${\bm I}_j(\theta) \cdot {\bm \chi}'(\theta) = 0$ by the biorthogonality of the left and right Floquet vectors on the limit cycle as explained in~\cite{Kuramoto2019,takata2021fast}. 
Therefore, the amplitude $r_j$ is of $O(\varepsilon)$ and always kept small because $\mbox{Re}\ \lambda_j < 0$, and the system state ${\bm X}$ remains near the limit cycle.

Thus, neglecting terms of $O(\varepsilon)$, we obtain a simple set of phase-amplitude equations approximately describing the $N$-dimensional system as
\begin{align}
\dot{\theta} = \omega + p(t), \quad \dot{r}_j = \lambda_j r_j, \quad (j=2, ..., N),
\end{align}
provided that the oscillator is in the close vicinity of the limit cycle and is driven only in the tangential direction to the limit cycle.
Note that the phase and amplitude sensitivity functions, ${\bm Z}$ and ${\bm I}_j$, are eventually not required to implement the tangential driving described above; only the phase $\theta$ and ${\bm \chi}'$ are necessary.
By retaining, for example, only the slowest decaying amplitude $r_1$, we can obtain a reduced set of phase-amplitude equations approximately.
The original system state can be approximately reconstructed as ${\bm X}(t) = {\bm \chi}(\theta(t))$ from the phase $\theta(t)$. 

The above procedure has also been extended to limit-cycle solutions of RD equations~\cite{nakao2014phase,nakao2021phase} and can also be formulated for fluid systems~\cite{kawamura2013collective,kawamura2015phase,taira2018phase,nair2021phase,nakao2021phase,loe2021phase,kawamura2022adjoint,godavarthi2023optimal}.
In this study, we consider a RD equation given by
\begin{align}
\label{eq:rd}
\frac{\partial {\bm X}({\bm x}, t)}{\partial t} = {\bm F}({\bm X}({\bm x}, t), {\bm x}) + D \frac{\partial^2 {\bm X}({\bm x}, t)}{\partial {\bm x}^2},
\end{align}
where ${\bm X}({\bm x}, t) \in {\mathbb R}^N$ is a field variable representing, for example, concentrations of $N$ chemical species at position ${\bm x} \in V$ and time $t \in {\mathbb R}$ in the domain $V$ of the $d$-dimensional space, 
${\bm F}$ represents the local dynamics of ${\bm X}$ at position ${\bm x}$ such as chemical reaction, 
$D \in {\mathbb R}^{N \times N}$ is a matrix of diffusion coefficients, and ${\partial}^2 / {\partial {\bm x}^2}$ represents the $d$-dimensional Laplacian operator.

We assume that the above RD equation has an exponentially stable limit-cycle solution ${\bm X}_0({\bm x}, t) = {\bm X}_0({\bm x}, t+T)$ representing a rhythmic spatiotemporal pattern.
We can then define an asymptotic phase functional $\Theta[ {\bm X}]$ that maps the field variable ${\bm X} \in B$ to $[0, 2\pi)$, and similarly  amplitude functionals $R_j[ {\bm X} ]$  ($j=1, 2, ...$) mapping ${\bm X} \in B$ to ${\mathbb C}$, satisfying
\begin{align}
\dot\Theta[{\bm X}] &= \int_V \frac{\partial {\bm X}({\bm x}, t)}{\partial t} \cdot \frac{\delta \Theta[{\bm X}]}{\delta {\bm X}({\bm x}, t)} d^d{\bm x} = \omega,
\cr
\dot R_j[{\bm X}] &= \int_V \frac{\partial {\bm X}({\bm x}, t)}{\partial t} \cdot \frac{\delta R_j[{\bm X}]}{\delta {\bm X}({\bm x}, t)} d^d{\bm x} = \lambda_j R_j[{\bm X}],
\end{align}
in the basin $B$ of the limit cycle, where $\lambda_j$ is the $j$th Floquet exponent of the limit-cycle solution.
Here, the relations in Eq.~(\ref{eq:phaseampode}) are generalized to the case where ${\bm X}$ is a field variable, and $\delta / \delta {\bm X}({\bm x}, t)$ represents a functional derivative acting on $\Theta$ or $R_j$.
Note that there can be infinitely many Floquet exponents for RD systems.
As before, we represent a system state on the limit cycle as ${\bm \chi}({\bm x}, \theta) = {\bm X}_0({\bm x}, \theta/\omega)$ as a function of $\theta = \Theta[{\bm X}]$, where $R_j[ {\bm \chi}(\cdot, \theta)] = 0$ on the limit cycle.

Assuming that the system state ${\bm X}$ is sufficiently close to the limit cycle, ${\bm X}({\bm x}, t) = {\bm \chi}({\bm x}, \theta(t)) + O(\varepsilon)$ with $\theta(t) = \Theta[ {\bm X}({\bm x}, t) ]$, and that the system state is driven only in the tangential direction ${\bm \chi}'({\bm x}, \theta) = \partial {\bm \chi}({\bm x}, \theta) / \partial \theta$ as
\begin{align}
\label{eq:rdperturb}
\frac{\partial {\bm X}({\bm x}, t)}{\partial t} = {\bm F}({\bm X}({\bm x}, t), {\bm x}) + D \frac{\partial^2 {\bm X}({\bm x}, t)}{\partial {\bm x}^2}
+ p(t) {\bm \chi}'({\bm x}, \theta(t)),
\end{align}
where $p(t)$ is the driving amplitude, we obtain a set of approximate phase-amplitude equations
\begin{align}
\dot{\theta} = \omega + p(t), \quad \dot{r}_j = \lambda_j r_j, \quad (j=1, 2, ... )
\end{align}
by neglecting small terms of $O(\varepsilon)$ in a similar manner to the previous case.
Thus, the phase-amplitude description can also be obtained for rhythmic RD systems exhibiting limit-cycle oscillations.
Although there can be infinitely many amplitudes, in practice, only the first several amplitudes are dominant while other  amplitudes decay quickly.
We can then reduce the dimensionality to several dimensions (one for the phase and several for the amplitudes) by focusing only on slowly decaying amplitudes.

We note that, as before, the phase and amplitude sensitivity functions are eventually not required for implementing the tangential driving described above.
This is indeed important for applying the phase autoencoder to spatiotemporal patterns because, even though the phase and amplitudes can be successfully learned by the phase autoencoder in the close vicinity of the limit cycle, accurate estimation of their gradients (i.e., phase and amplitude sensitivity functions) can be quite challenging due to numerical errors caused by high-dimensionality. 
With the above method, only the information along the limit cycle (i.e., the tangential direction) is necessary, which can be estimated from data stably in contrast to the gradients.

\section{Phase autoencoder}

\subsection{Phase autoencoder for ordinary limit cycles}

The simplicity of the phase equation has facilitated detailed analysis of the synchronization dynamics of coupled oscillators~\cite{Kuramoto1984chemical,Hoppensteadt1997weakly,Brown2004phase,Ermentrout2010mathematical,Nakao2016phase}. However, when the mathematical model of the system is not available,
the asymptotic phase function (or its gradient) need to
be inferred in a data-driven manner. Various methods for inferring the phase dynamics have been discussed in the literature~\cite{matsuki2023extended,rosenblum2023inferring,furukawa2024bayesian,namura2022estimating,yamamoto2025gaussian,takata2023definition}. In this study, we focus on and extend the {\it phase autoencoder} proposed in~\cite{yawata2024phase}, which is 
a physics-informed neural network based on the autoencoder architecture.

The autoencoder~\cite{hinton2006reducing} is a self-supervised neural network used for dimensionality reduction and feature extraction, consisting of an encoder, which maps the input data to a low-dimensional latent-space variable and extracts low-dimensional features of the input data, and a decoder, which reconstructs the original data from the low-dimensional latent-space variable. It has been widely used in applications including classification~\cite{ghasedi2017deep}, anomaly detection~\cite{gong2019memorizing}, and noise reduction~\cite{eraslan2019single}.
Recently, studies of dynamical systems based on autoencoders have been conducted~\cite{lusch2018deep,fukami2023grasping,bakarji2023discovering,yawata2024phase}. 
In these methods, the dynamical information of the system is used as physical constraints on the latent-space variables to correctly extract the essential dynamical features. 

We denote the encoder and decoder of the autoencoder as $f_{enc} : S \to L$ and $f_{dec} : L \to S$, respectively,
where $S$ is a $D_S$-dimensional space of input data and $L$ is a $D_L$-dimensional latent space. The dimension $D_S$ of the input space is usually large and dimension $D_L$ of the latent space is small. The high-dimensional input data ${\bm X} \in S$ is transformed to a low-dimensional latent variable ${\bm Y} = f_{enc}({\bm X}) \in L$ by the encoder, and then the decoder gives an approximate reconstruction $\hat{\bm X} = f_{dec}({\bm Y}) \in S$ of the original data ${\bm X}$ as closely as possible.

In the phase autoencoder~\cite{yawata2024phase}, the state of a limit-cycle oscillator ${\bm X}_t \in {\mathbb R}^{D_S}$ at discrete time $t$ is transformed to a latent variable ${\bm Y}_t = (Y_{0,t}, Y_{1,t}, Y_{2,t},...Y_{D_L-1,t}) \in {\mathbb R}^{D_L}$ to satisfy the following conditions: 
\begin{equation}
\label{eq:evolve}
\begin{aligned}
&\quad \quad \quad Y_{0,t}^2 + Y_{1,t}^2 = 1, \\
Y_{0,t+\Delta t} &= Y_{0,t} \cos \omega \Delta t - Y_{1,t} \sin \omega \Delta t, \\
Y_{1,t+\Delta t} &= Y_{0,t} \sin \omega \Delta t + Y_{1,t} \cos \omega \Delta t, \\
Y_{j, t + \Delta t} &= Y_{j, t} \exp( \lambda_j \Delta t ), \quad (j = 2, \dots, D_L),
\end{aligned}
\end{equation}
where $\Delta t$ is the step interval of the input data, $\omega$ is the frequency, and $\lambda_j < 0$ is the $(j+1)$th Floquet exponent of the limit cycle.
In~\cite{yawata2024phase}, $D_L$ was fixed to $3$, so the latent variable had only $3$ components. In this study, to learn high-dimensional data, we extend the dimension of the latent space $D_L$ to an arbitrary value up to $D_S$ ($D_S \leq N$).
Among the components of the latent variable ${\bm Y}$, $Y_{0,t}$ and $Y_{1,t}$ directly represents the asymptotic phase, and $Y_{3,t}$, ..., $Y_{D_L}$  represent the remaining amplitude components that decay exponentially with different decay rates $\lambda_3, ..., \lambda_{DL}$.
In the training, the natural frequency and Floquet exponents are also estimated from the input data in addition to the autoencoder's parameters.

By using the trained phase autoencoder, the phase function $\Theta({\bm X})$ of the oscillator is obtained as 
\begin{align}
\Theta({\bm X}) = \arctan \left( \frac{ Y_{1} }{ Y_{0} } \right) = \arctan \left( \frac{ f_{1,enc}({\bm X}) }{ f_{0,enc}({\bm X})} \right),
\end{align}
and  the amplitudes as
\begin{align}
R_j( {\bm X} ) = f_{j, enc}({\bm X}),
\end{align}
where $f_{k, enc}({\bm X})$ represents the $k$th component of the vector-valued function $f_{enc}({\bm X})$.

\begin{figure*}[htbp]
\includegraphics[width=0.7\hsize]{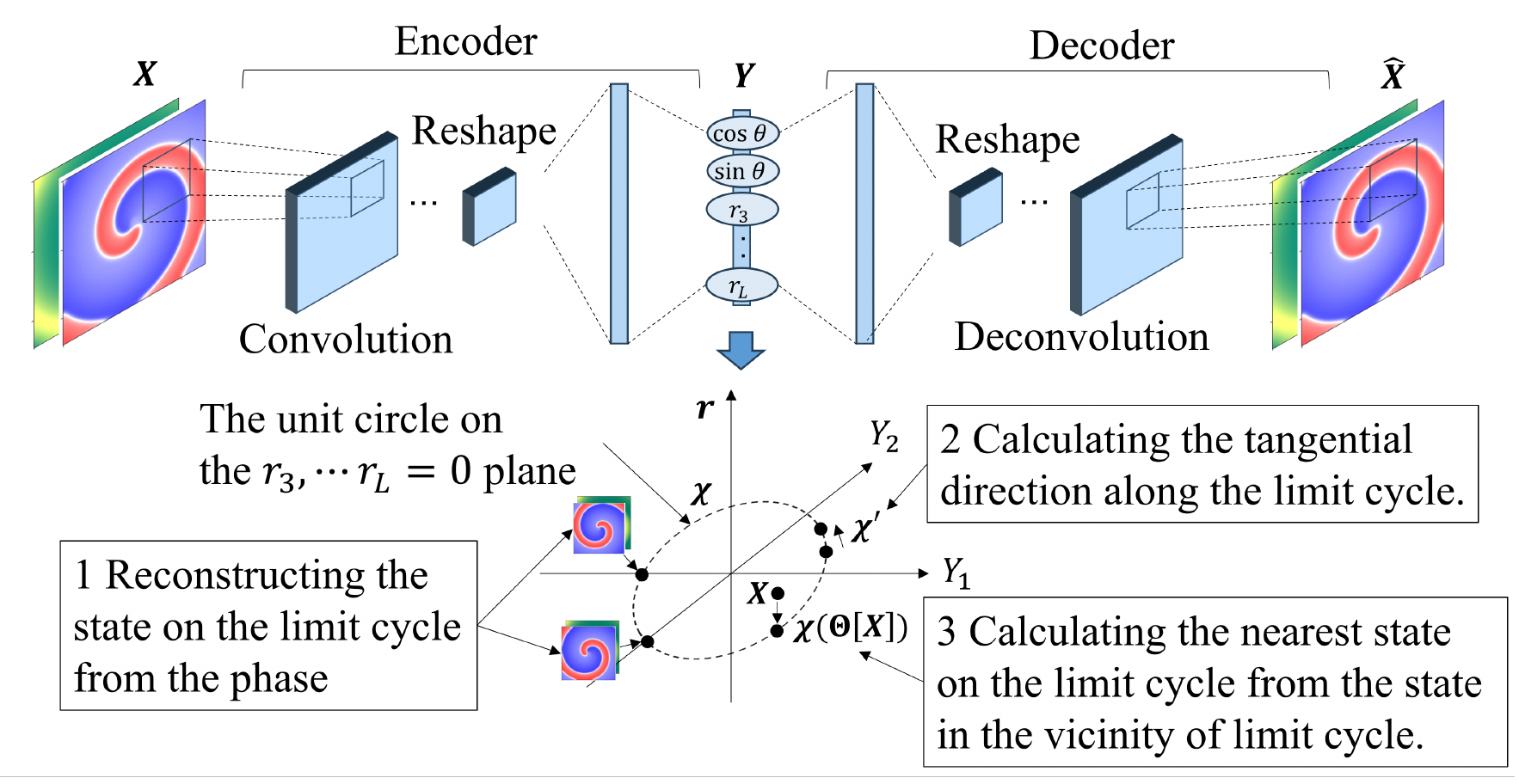}
\caption{Phase autoencoder for rhythmic spatiotemporal patterns. By extending the phase autoencoder with convolutional layers, we enable phase estimation, computation of the tangential direction along the limit cycle, and identification of the state on the limit cycle corresponding to a given system state (pattern).}
\label{fig1}
\end{figure*}
\subsection{Phase autoencoder for rhythmic spatiotemporal patterns}
The examples of limit cycles considered in~\cite{yawata2024phase} were low-dimensional, such as the FitzHugh-Nagumo model with two variables or Hodgkin-Huxley model with four variables. 
However, the phase autoencoder framework is expected to be applicable to high-dimensional spatiotemporal patterns, 
provided that some extension to handle spatiotemporal patterns is introduced.
In this study, we extend the phase autoencoder to extract phase dynamics from high-dimensional data of rhythmic spatiotemporal patterns in RD systems by introducing three key improvements (Fig.~{\ref{fig1}}).

First, while the original phase autoencoder employed a fully connected neural network, we introduce convolutional layers~\cite{lecun1989handwritten,krizhevsky2012imagenet}, enabling the model to  effectively capture spatial and sequential features in both 2D and 1D patterns while reducing the number of learnable parameters.
This reduction in parameters leads to lower computational costs during both training and inference.
Second, we increase the dimension $D_L$ of latent variables in the autoencoder, which enhances its representational capacity around the limit cycle and improves training stability.
Third, we add the Gaussian noise to the input data during training, which induces inputs that are close in the Euclidean space of the system states to be projected into nearby regions in the latent space, thereby stabilizing phase estimation. 

By using the above framework illustrated in Fig.~{\ref{fig1}}, we train a phase autoencoder for rhythmic RD systems, which gives the asymptotic phase $\theta(t)$ of the (discretized) input system state (i.e., pattern) ${\bm X}({\bm x}, t)$ as 
\begin{align}
\theta = \Theta[ {\bm X} ] = \arctan \left( \frac{ f_{1,enc}[{\bm X}] }{ f_{0,enc}[{\bm X}]} \right),
\end{align}
and the amplitude as
\begin{align}
r_j = R_j[ {\bm X} ] = f_{j, enc}[{\bm X}],
\end{align}
where the field variable ${\bm X}$ is appropriately discretized when given to the autoencoder.
The original spatiotemporal pattern with phase $\theta$ is approximately reconstructed as
\begin{align}
\hat{\bm X}({\bm x}, \theta) = f_{dec}( {\bm x}, \theta ).
\end{align}
We briefly describe the training procedure of the phase autoencoder using the case of $D_L=3$ as an example.
In training, $f_{enc}$ and $f_{dec}$ are optimized so that the latent variable ${\bm Y}_t$ satisfies Eq.~(\ref{eq:evolve}).
One principal loss function is the reconstruction loss:
\begin{gather}
    L_{recon} = {\mathbb E} \left[ \left\|{\bm X}_t - f_{dec}(f_{enc}({\bm X}_t)) \right\|^2 \right],
    \label{eq:reconloss}
\end{gather}
where $\| \cdot \|$ represents the Euclidean norm, ${\bm X}_t$ represents the input data, and $\mathbb{E}$ represents the expectation with respect to the input data. 
In addition to the reconstruction loss, the latent variable is expressed as a three-dimensional vector ${\bm Y}_t=[Y_{0,t}\;Y_{1,t}\;Y_{2,t}]^\top=f_{enc}({\bm X}_t)$, and the dynamical loss $L_{dyn}$, another principal loss function, is used to constrain the time evolution in the latent space.
Here, $Y_{0,t}$ and $Y_{1,t}$ are normalized so that $Y_{0,t}^2+Y_{1,t}^2=1$ in the encoding process.
The dynamical loss is:
\begin{gather}
    \footnotesize
     L_{dyn} = {\mathbb E} \left[ \left\|{\bm Y}_{t+\Delta t} -\begin{bmatrix}
    \cos \omega \Delta t & -\sin \omega \Delta t & 0\\
     \sin \omega \Delta t & \cos \omega \Delta t & 0\\
    0 & 0 & e^{-\lambda\Delta t}
    \end{bmatrix}{\bm Y}_t \right\|^2 \right].
\end{gather}
This loss function requires that $Y_{0,t}$ and $Y_{1,t}$ evolve in the $Y_{0}$-$Y_{1}$ plane at a constant angular velocity $\omega$, and $Y_{2,t}$ approaches 0 asymptotically over time.
In this way, the unit circle 
in the $Y_{0}$-$Y_{1}$ plane corresponds to the limit cycle, 
and the two latent variables $Y_{0,t}, Y_{1,t}$ represent the asymptotic phase
as $\theta_t=\text{arctan}(Y_{0,t}, Y_{1,t})$.
Thus, the encoder $f_{enc}$ gives an approximation of the asymptotic phase function.

\subsection{Synchronization aided by phase autoencoder}

We consider the problem of designing a driving input for the RD system, Eq.~(\ref{eq:rd}), to achieve desired phase dynamics, for example, synchronization of rhythmic spatiotemporal patterns. This can be accomplished in a data-driven manner using the phase autoencoder.

Suppose that we drive the state ${\bm X}$ of the RD system near the limit cycle ${\bm \chi}$  in the tangential direction ${\bm \chi}' = \partial {\bm \chi} / \partial \theta$ as
\begin{align}
\label{eq:rdcontrol}
\frac{\partial {\bm X}}{\partial t} = {\bm F}({\bm X}, {\bm x}) + D \frac{\partial^2 {\bm X}}{\partial {\bm x}^2}  + g(\theta, t) {\bm \chi}'
({\bm x}, \theta),
\end{align}
where $\theta = \Theta[ {\bm X} ]$ is the phase of the system and $g(\theta, t)$ is a given function.
From Eq.~(\ref{eq:rdperturb}), the phase $\theta$ and each amplitude $r_j = R_j[ {\bm X} ]$ approximately obey
\begin{align}
\dot{\theta} = \omega + g(\theta, t),
\quad
\dot{r}_j = \lambda_j r_j, 
\quad
(j=1, 2, ...).
\end{align}
Thus, by driving the RD system in the tangential direction, we can control the phase dynamics in the latent space while keeping the amplitude dynamics unaffected. 

In practice, the driving input can also affect the amplitude dynamics $r_j$ due to numerical discretization and other factors.
To suppress the amplitudes and keep the system state close to the limit cycle also in such cases, we can further introduce a feedback control to
Eq.~\eqref{eq:rdcontrol} as
\begin{align}
\label{eq:rdcontrolfeedback}
\frac{\partial {\bm X}}{\partial t} = {\bm F}({\bm X}, {\bm x}) + D \frac{\partial^2 {\bm X}}{\partial {\bm x}^2}  +  g(\theta, t) {\bm \chi}'
({\bm x}, \theta)
\cr
 - k \{ {\bm X}({\bm x}, t) - {\bm \chi}({\bm x}, \theta) \},
\end{align}
where $k \geq 0$ is the feedback gain. 
This additional feedback suppresses the amplitude deviations of the actual system state ${\bm X}$ from the corresponding state ${\bm \chi}$ on the limit cycle with the phase value $\theta = \Theta[{\bm X}]$, thereby stabilizing the control of the phase dynamics.

We emphasize that, in the phase autoencoder, we can estimate the phase $\theta$ from the system state ${\bm X}$ using the encoder as $\theta = f_{enc}({\bm X})$, and obtain the tangential direction ${\bm \chi}'({\bm x}, \theta)$ from $\theta$ as $\partial f_{dec}({\bm x}, \theta) / \partial \theta$ using the decoder (see Fig.~\ref{fig1}).
This is the advantage of the phase autoencoder; since we can approximately reconstruct the original system state, we can drive the system in the tangential direction in a data-driven manner, enabling data-driven phase control of limit cycles.
Below, we introduce two methods, a reference-based method and a coupling-based method, for synchronizing multiple RD systems aided by the phase autoencoder.

In the reference-based method (entrainment), we entrain the RD systems to a reference phase $\theta_r$
that evolves at a constant frequency $\omega_r$. In this case, we drive the RD system as
\begin{align}
\frac{\partial {\bm X}}{\partial t} &= {\bm F}({\bm X}, {\bm x}) + D \frac{\partial^2 {\bm X}}{\partial {\bm x}^2}  +  \eta\sin (\theta_r - \theta) {\bm \chi}'({\bm x}, \theta),
\label{eq:ref_con}
\end{align}
where $\eta$ is the intensity of the driving input, $\theta_r = \omega_r t$ is the reference phase, $\theta = \Theta[ {\bm X} ]$ is the phase of the system state, and $\omega_r$ is the driving frequency close to the natural frequency $\omega$.
The reduced phase equation is given by
\begin{align}
\dot{\theta} = \omega +  \eta\sin (\omega_r t - \theta),
\end{align}
and introducing the phase difference $\psi = \theta - \omega_r t$, we obtain
\begin{align}
\dot{\psi} = (\omega - \omega_r) - \eta\sin \psi.
\end{align}
If $| \left(\omega_r - \omega\right)/\eta| < 1$, this system has a stable fixed point 
$\psi_0 = \sin^{-1}\left(\left(\omega_r - \omega\right)/\eta\right)$.
Thus,  $\theta = \omega_r t + \psi_0$, i.e., the system is entrained to the periodic forcing by adjusting its frequency to $\omega_r$.

In coupling-based method (mutual synchronization), we couple two RD systems as
\begin{align}
\label{eq:aemutual}
\frac{\partial {\bm X}_1}{\partial t} = {\bm F}_1({\bm X}_1, {\bm x}) + D_1 \frac{\partial^2 {\bm X}_1}{\partial {\bm x}^2}  +  \eta\sin (\theta_2 - \theta_1) {\bm \chi}_1'({\bm x}, \theta_1),
\cr
\frac{\partial {\bm X}_2}{\partial t} = {\bm F}_2({\bm X}_2, {\bm x}) + D_2 \frac{\partial^2 {\bm X}_2}{\partial {\bm x}^2}  +  \eta\sin (\theta_1 - \theta_2) {\bm \chi}_2'({\bm x}, \theta_2),
\end{align}
where ${\bm X}_1$, ${\bm X}_2$ are the field variables of the two rhythmic RD systems, ${\bm F}_1, {\bm F}_2$ are their local dynamics, $D_1, D_2$ are the diffusion matrices, ${\bm \chi}_1$ and ${\bm \chi}_2$ are stable limit-cycle solutions, and $\theta_1 = \Theta[ {\bm X}_1 ]$, $\theta_2 = \Theta[ {\bm X}_2 ]$ are their phase values , respectively.
That is, the encoder $f_{enc}$ is used to compute the phase of each RD system, and the decoder $f_{dec}$ is used to calculate the tangential direction of each RD system, which is then used to determine the control input.
The reduced phase equations are
\begin{align}
\dot{\theta}_1 &= \omega_1 +  \eta\sin (\theta_2 - \theta_1), \cr
\dot{\theta}_2 &= \omega_2 +  \eta\sin (\theta_1 - \theta_2),
\end{align}
and, introducing the phase difference $\psi = \theta_1 - \theta_2$, we obtain
\begin{align}
\dot{\psi} = (\omega_1 - \omega_2) -  2 \eta\sin \psi,
\end{align}
where $\omega_1$ and $\omega_2$ are the natural frequencies of the two systems.
Thus, if $|(\omega_1 - \omega_2)/\eta| < 2$, the two RD systems synchronize with each other.

Note that, in both methods, we assumed the simplest sinusoidal functions for $g(\theta, t)$ as examples, but $g(\theta, t)$ can be chosen arbitrary.
We may also use, for example, the optimized coupling functions~\cite{Monga2019optimal,takata2021fast,kato2021optimization,wilson2022recent,godavarthi2023optimal,mircheski2023phase,namura2024optimal} for more efficient synchronization.

\section{Examples}

\subsection{FitzHugh-Nagumo reaction-diffusion system}

As examples of rhythmic spatiotemporal patterns, we analyze oscillating spots and spiral waves in the FitzHugh-Nagumo RD system of excitable chemical reactions. The model is given by Eq.~(\ref{eq:rd}), where
\begin{align}
{\bm X}({\bm x}, t) = \begin{pmatrix} u({\bm x}, t) \\ v({\bm x}, t) \end{pmatrix},
\end{align}
and
\begin{align}
{\bm F}(u, v) = \begin{pmatrix} u (u-\alpha) (1-u) - v \\ \tau^{-1} ( u - \gamma v ) \end{pmatrix},
\quad
D = \begin{pmatrix} \kappa & 0 \\ 0 & \delta \end{pmatrix},
\end{align}
where $u({\bm x}, t)$ and $v({\bm x}, t)$ are the activator and inhibitor concentrations at position ${\bm x}$ and time $t$, $\alpha$ and $\gamma$ are parameters, $\tau$ is a time constant, 
and $\kappa$ and $\delta$ are the diffusion coefficients of $u$ and $v$, respectively.
See~\cite{nakao2014phase} for the details of the numerical simulations as well as the theoretical, model-based phase-reduction analysis of this RD system.

\subsection{1D oscillating spots}

\begin{figure}[htbp]
\includegraphics[width=\hsize]{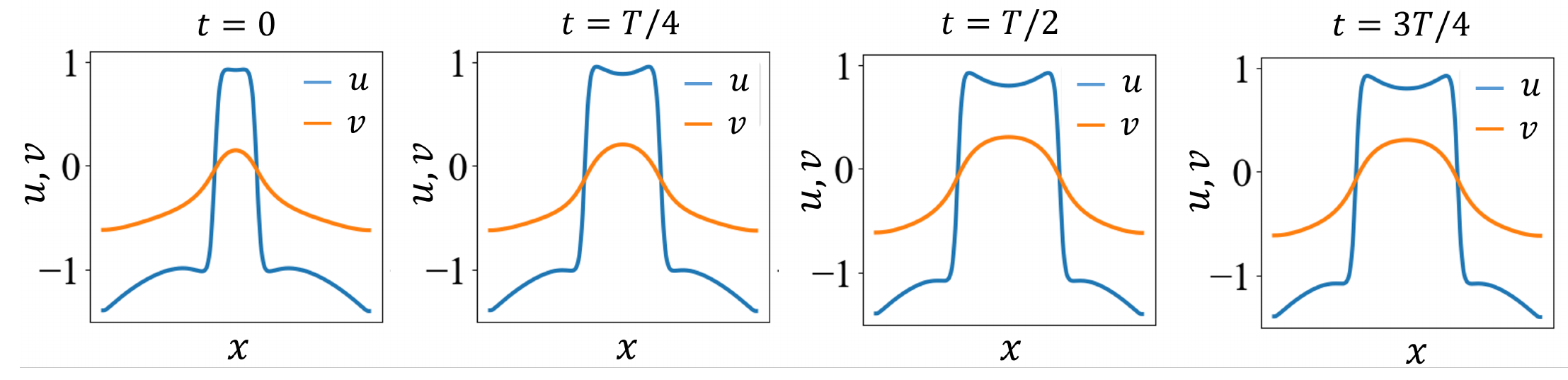}
\caption{Oscillating spot patterns. Time evolution of the field variables on the limit cycle of the RD system. The blue lines indicate the $u$ component and the orange lines indicate the $v$ component as functions of position $x$ at time $T=0, T/4, 2T/4$, and $3T/4$.}
\label{fig2a}
\end{figure}

\begin{figure}[htbp]
\includegraphics[width=\hsize]{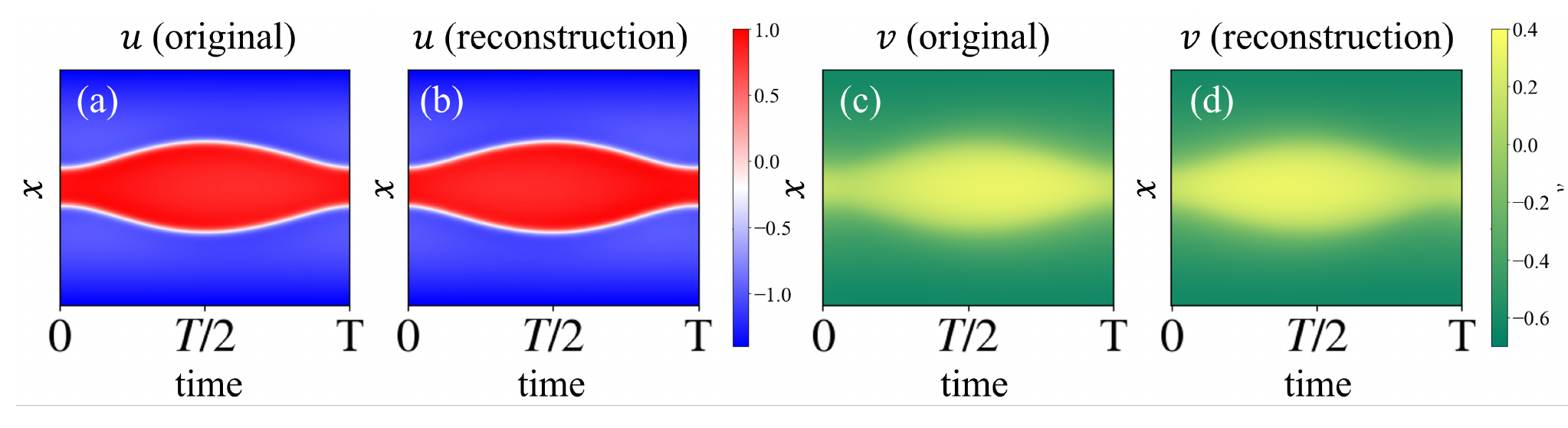}
\caption{Learning of oscillating spot patterns. (a,c) Original spatiotemporal patterns on the limit cycle. (b,d) Reconstructed spatiotemporal patterns on the limit cycle. In each panel, the time evolution of the $u$ component is shown in a density plot for one period of oscillation ($0 \leq \theta < 2\pi$).}
\label{fig:spot_recon}
\end{figure}

For the oscillating spots, we consider a one-dimensional system of length $L=80$ with no-flux boundary conditions, and assume $\alpha(x) = \alpha_0 + ( \alpha_1 - \alpha_0 ) ( 2 x / L - 1 )^2$ with $\alpha_0 = -1.1$ and $\alpha_1 = -1.6$, where $0 \leq x \leq L$ is the position.  Other parameters are $\tau^{-1} = 0.03$, $\gamma = 2.0$, $\kappa = 1$, and $\delta = 2.5$. 
In the numerical simulations, we use $m=320$ grid points to discretize each of the one-dimensional fields $u$ and $v$, 
yielding a system of $640$ dimensions.
With these conditions, an oscillating spot pinned at the center ($x=L/2$) can be generated as shown in Fig.~\ref{fig2a} and Figs.~\ref{fig:spot_recon}(a) and (c). The natural period of the oscillation is $T \approx 196.5$.

The phase autoencoder takes a matrix input of the form 
$({\bm X}_0, ..., {\bm X}_{m-1}) \in \mathbb{R}^{2 \times m}$, discretizing the original field variable ${\bm X}(x)$ using $m$ grid points as ${\bm X}_s = {\bm X}(x=L s / m)$ for $s=0, ..., m-1$. This is suitable for one-dimensional convolution, which stabilizes training and reducing computational costs.
In the training, we set the number of initial system states to $n = 100$ and carried out numerical simulations for a duration of $3T$ starting from these initial states.
The initial states are chosen randomly using the standard deviation and the mean of each component $u$ or $v$ on the limit cycle,
\begin{align}
    \sigma_u &= \sqrt{\frac{1}{2\pi L}\int_0^{2\pi}\int_0^L \left||{\bm \chi}_u({x}, \theta)-\langle{\bm \chi}_u\rangle\right||^2dxd\theta},\\
    \sigma_v &= \sqrt{\frac{1}{2\pi L}\int_0^{2\pi}\int_0^L \left||{\bm \chi}_v({x}, \theta)-\langle{\bm \chi}_v\rangle\right||^2dxd\theta},\\
    \langle{\bm \chi}_u\rangle &= \frac{1}{2\pi L}\int_0^{2\pi}\int_0^L {\bm \chi}_u({x}, \theta)dxd\theta,\\
    \langle{\bm \chi}_v\rangle &= \frac{1}{2\pi L}\int_0^{2\pi}\int_0^L {\bm \chi}_v({x}, \theta)dxd\theta.
\end{align}
Using these values, we generate $n$ discretized initial states $({\bm X}_{1}^{(j)}, ..., {\bm X}_{m}^{(j)})$ ($j=0, ..., n-1$) as
\begin{align}
    {\bm X}_{s}^{(j)} = {\bm \chi}_{s}\left(\frac{2\pi j}{n} \right) + \beta_1
    \begin{bmatrix}
\sigma_u & 0 \\
0 & \sigma_v
\end{bmatrix}
{\bm \Xi}_1,
\quad
(s=0, ..., m-1),
\end{align}
where $\left( {\bm \chi}_1\left(\frac{2\pi j}{n}\right), ..., {\bm \chi}_m\left(\frac{2\pi j}{n})\right)\right)$ is the discretized field variable of the system state ${\bm \chi}({\bm x}, \theta)$ at $\theta = 2 \pi j / n$ on the limit cycle,
${\bm \Xi}_1 \in \mathbb{R}^{2 \times 320}$ is a random matrix whose element is independently drawn from a normal distribution, ${\bm \Xi}_{1,ij} \sim \mathcal{N}(0, 1)$,
and the parameter $\beta_1$ represents the magnitude of the variability in the initial values, which is set to $0.5$ in this study. 
In addition, during training, we added observation noise at each discrete time step as 
\begin{align}
    {\bm X}_{s}^{(obs)} = {\bm X}_s + \beta_2
    \begin{bmatrix}
\sigma_u & 0 \\
0 & \sigma_v
\end{bmatrix}
{\bm \Xi}_2,
\quad
(s=0, ..., m-1),
\end{align}
where ${\bm \Xi}_2 \in \mathbb{R}^{2 \times 320}$ is a random matrix with each element obeying ${\bm \Xi}_{2,ij} \sim \mathcal{N}(0, 1)$, and $\beta_2$ is a parameter that controls the magnitude of the observation noise. We set $\beta_2 = 0.01$ in this study.
By adding observation noise, phase estimation in the vicinity of the limit cycle is stabilized.
We chose the dimension of the latent space as $D_L = 5$ and the step interval as $\Delta t = 10.0$. 
The dimension of the latent space is increased compared to $D_L=3$ in~\cite{yawata2024phase} to stabilize the learning of high-dimensional systems.

We trained the phase autoencoder using the spatiotemporal data obtained by numerical simulations of Eq.~(\ref{eq:rd}) and verified that it can assign appropriate phase values to the input patterns and approximately reconstruct the original patterns from the phase values. 
In Figs.~\ref{fig:spot_recon}, using the trained phase autoencoder, the original patterns (input) and the reconstructed patterns (output) are compared for one period of oscillation, showing good agreement. 

\begin{figure}[htbp]
\includegraphics[width=\hsize]{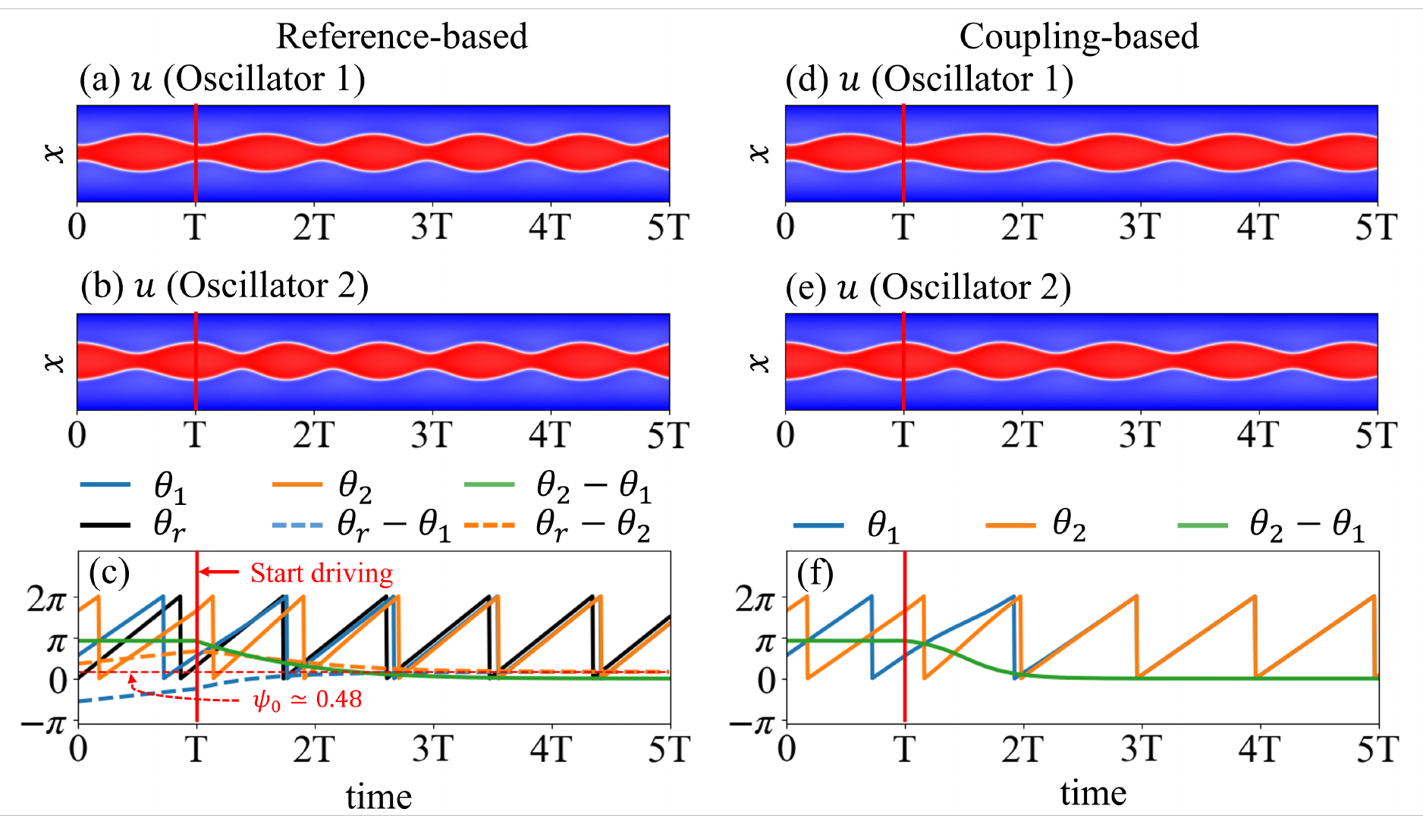}
\caption{Autoencoder-aided rapid synchronization of two oscillating spots.
(a-c) Reference-based method. Panels (a) and (b) show the evolution of the field variable $u$ for RD systems $1$ and $2$, respectively, and panel~(c) shows the reference phase $\theta_r$, phases $\theta_1$ and $\theta_2$ of the two RD systems, and their phase differences $\theta_r - \theta_1$, $\theta_r - \theta_2$, and $\theta_2 - \theta_1$.
(d-f) Coupling-based method. Panels (d) and (e) show the evolution of $u$ for two RD systems and panel (f) shows the phases $\theta_1$ and $\theta_2$ of the two RD systems and their difference $\theta_2 - \theta_1$. 
The vertical red lines indicate the time at which the driving or coupling is introduced.
The color scale is the same as in Fig.~\ref{fig:spot_recon}.
}
\label{fig:spot_control}
\end{figure}

First, we demonstrate the results of the reference-based method described by Eq.~(\ref{eq:ref_con}) for synchronizing two RD systems with identical properties.
We chose two states with different phase values on the limit cycle as initial conditions for the two systems, evolve them from $0$ to $T$ without the driving input, and applied the driving input from $t=T$.
The reference frequency is set to $\omega_r = 1.15 \times \omega$, which is larger than the natural frequency $\omega$ of the two oscillating spots, and the driving intensity $\eta$ is set to $0.01$.

Figures~\ref{fig:spot_control}(a) and (b) show the time evolution of the field variable $u$ of two RD systems $1$ and $2$, respectively.
Despite different initial states, the two RD systems eventually synchronize.
Figure~\ref{fig:spot_control}(c) shows the reference phase $\theta_r$ and the phases $\theta_1$ and $\theta_2$ of the two RD systems,
as well as their phase differences.
Since $\theta_r$ evolves at a higher frequency $\omega_r$ than the natural frequency $\omega$ of the RD systems,
the phase difference $\theta_r - \theta_1$ between the reference and the leading RD system 1 decreases,
while $\theta_r - \theta_2$ between the reference and the lagging RD system 2 increases
during $0 \leq t < T$.
As the driving input is applied from $t=T$, the phase differences $\theta_r - \theta_1$ and $\theta_r - \theta_2$ gradually converge to a common value, $\psi_0 \simeq 0.48$, indicating that the two RD systems are synchronized. Note that a phase lag $\theta_r - \theta_{1,2}$ remains finite because  $\omega_r > \omega$.

Next, we demonstrate the result of synchronizing two RD systems using the coupling-based method described by Eq.~(\ref{eq:aemutual}).
Figures~\ref{fig:spot_control}(d) and (e) show the spatiotemporal patterns of RD systems 1 and 2, respectively,
and Fig.~\ref{fig:spot_control}(f) shows the phases $\theta_1$ and $\theta_2$ of the RD systems 1 and 2 as well as their phase difference $\theta_1 - \theta_2$.
As before, the coupling is introduced from $t = T$ with $\varepsilon = 0.01$.
We can confirm that the two RD systems are mutually synchronized by the effect of the coupling input.

Noteworthy here is that both methods achieve nearly complete synchronization of the two RD systems within two or three oscillation periods, despite relying on phase reduction. 
This is because relatively large driving or coupling intensities can be used, as the driving input is applied only in the tangential direction along the limit cycle, avoiding amplitude deviations that would break down the phase description.
This is made possible by the phase autoencoder, which can reconstruct the spatiotemporal pattern via the decoder and evaluate the tangential direction in a data-driven manner.

Finally, we present the results of the method with feedback control for suppressing amplitude variations, as described by Eq.~(\ref{eq:rdcontrolfeedback}).
In the previous simulations, the phases were estimated and the driving inputs were updated at every time step $\Delta t_c$ of the numerical integration (i.e., $\Delta t_c = dt = 0.01$, where $dt$ is the integration time step).
However, in real control applications, $\Delta t_c$  may take large values, and errors in the amplitude direction can become more pronounced due to inaccurate estimation of the tangential direction.
In such situations, amplitude feedback becomes important.
%
In the following simulation, the driving input is updated less frequently, at much larger intervals of $\Delta t_c = 5.0$ compared to $dt = 0.01$, to reflect more realistic experimental conditions.
As a result, the direction of the control input deviates from the tangential direction along the limit cycle. 
We set the coupling intensity to $\eta = 0.185$ and the feedback gain to $k = 0.1$.

\begin{figure}[htbp]
\includegraphics[width=\hsize]{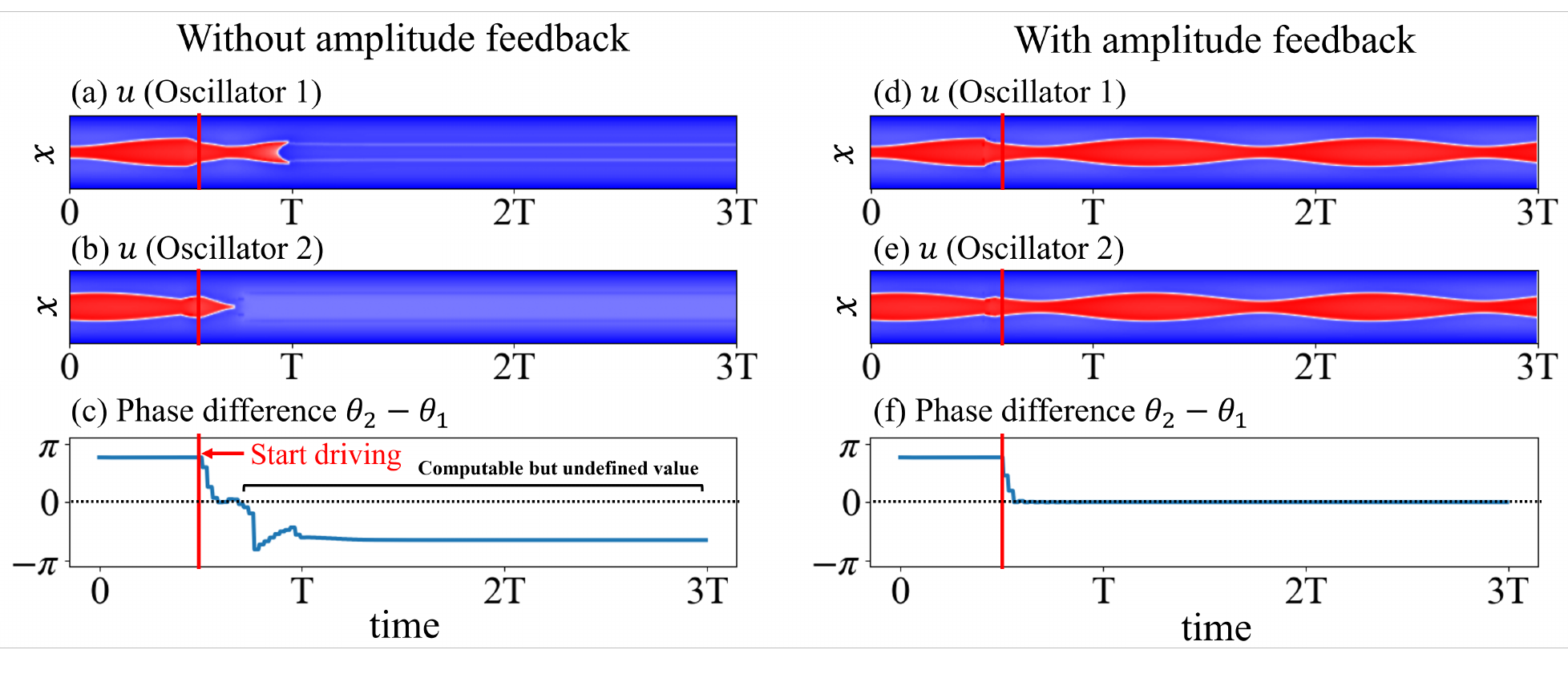}
\caption{Effect of amplitude feedback on coupling-based phase synchronization.
(a-c) Without amplitude feedback. Panels (a) and (b) show the evolution of $u$ for systems $1$ and $2$, respectively,
and panel~(c) shows the evolution of the phase difference.
Note that, although the phase autoencoder computes the phase values at any moment, the phase values are undefined after the collapse of the oscillating spots.
(d-f) With amplitude feedback. Panels (d) and (e) show the evolution of $u$ for systems $1$ and $2$, respectively, and panel~(f) shows the evolution of the phase difference.
The color scale is the same as in Fig.~\ref{fig:spot_recon}.
}
\label{fig:spot_sync4}
\end{figure}

Figure~\ref{fig:spot_sync4} compares the results of the coupling-based method with and without amplitude feedback.
In both cases, the coupling input is applied from $t = 0.5T$.
Figures~\ref{fig:spot_sync4}(a)-(c) show the case without amplitude feedback, while (d)-(f) show the case with amplitude feedback;
the first and second rows depict the evolution of the field variable $u$, respectively, and the third row shows the phase difference
between the two systems in each case.
In the absence of amplitude feedback, both RD systems exhibit noticeable deviations from the limit cycle, causing the oscillating spots to disappear. In contrast, when the amplitude feedback is introduced, the oscillating spots are preserved, and the two RD systems asymptotically synchronize with each other.


\subsection{2D spiral waves}

\begin{figure*}[htbp]
\includegraphics[width=\hsize]{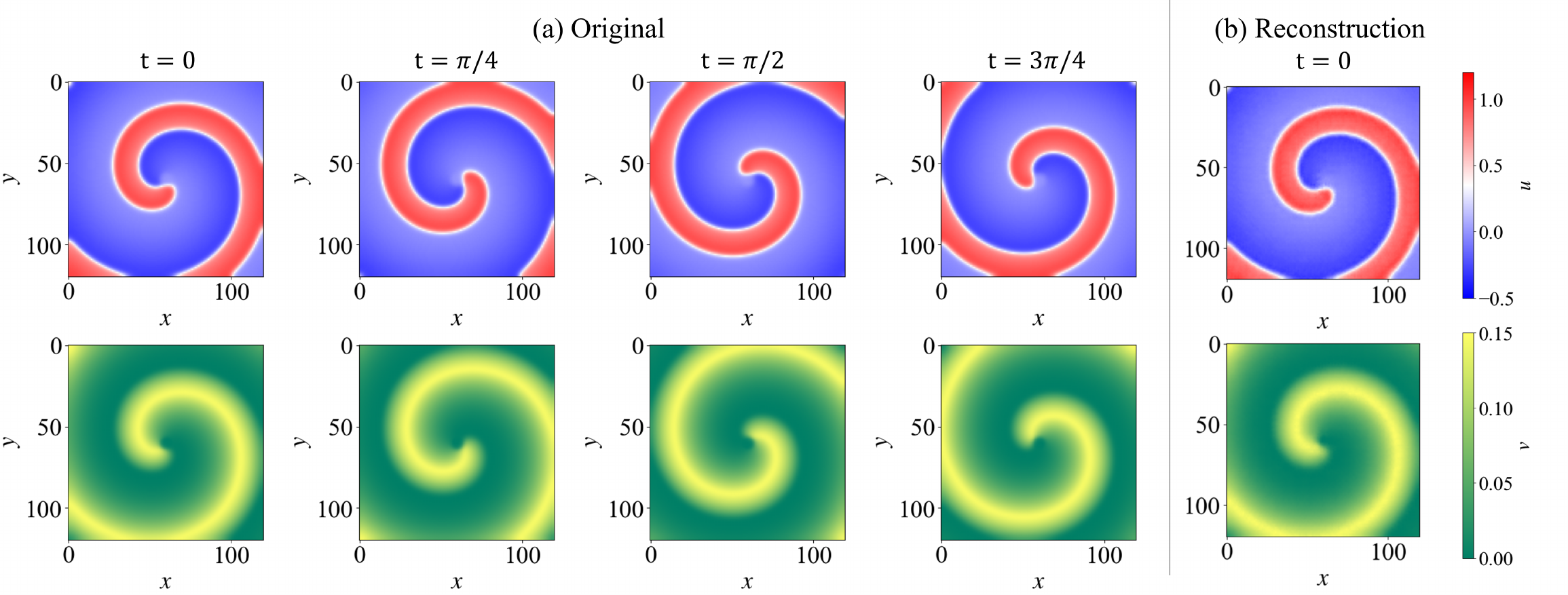}
\caption{Spiral waves patterns.
(a) Time evolution of the field variables on the limit cycle of the RD system.
The top panels show the $u$ component and the bottom panels show the $v$ component as functions of $x$ and $y$ at time $t=0, T/4, T/2$ and $ 3T/4$.
(b) Reconstructed spatiotemporal patterns on the limit cycle at $t=0$.
}
\label{fig:spiral_data}
\end{figure*}

For the spiral waves, we consider a two-dimensional square domain of side length $L=120$ with no-flux boundary conditions,
and assume $\alpha(x, y) = \alpha_0 + ( \alpha_1 - \alpha_0 ) \exp ( - r^4 / r_0^4 )$ with $\alpha_0 = 0.05$ and $\alpha_1 = 0.5$,
where $0 \leq x, y \leq L$ represents the position and $r = [ ( x - L/2 )^2 + ( y - L/2 )^2 ]^{1/2}$ is the distance from the center.
Other parameters are $\tau^{-1} = 0.005$, $\gamma = 2.5$, $\kappa = 0.4$, and $\delta = 0$. With these conditions, a stable spiral wave
pinned at the center can be generated as shown in Fig.~\ref{fig:spiral_data}(a), whose natural period is $T \approx 171.3$.
In the training, we set the number of initial states as $n = 100$,
the dimension of the latent space as $D_L = 20$, and the step interval as $\tau = 9.0$.
The other parameters are identical to those used for the oscillating spot.
Figure~\ref{fig:spiral_data}(a) shows typical one-period evolution of the field variables $u$ and $v$ on the limit cycle, and Fig.~\ref{fig:spiral_data}(b) shows the reconstructed field variables at $t=0$ using the trained decoder. Both $u$ and $v$ are accurately reconstructed.
 
%

\begin{figure}[htbp]
\includegraphics[width=\hsize]{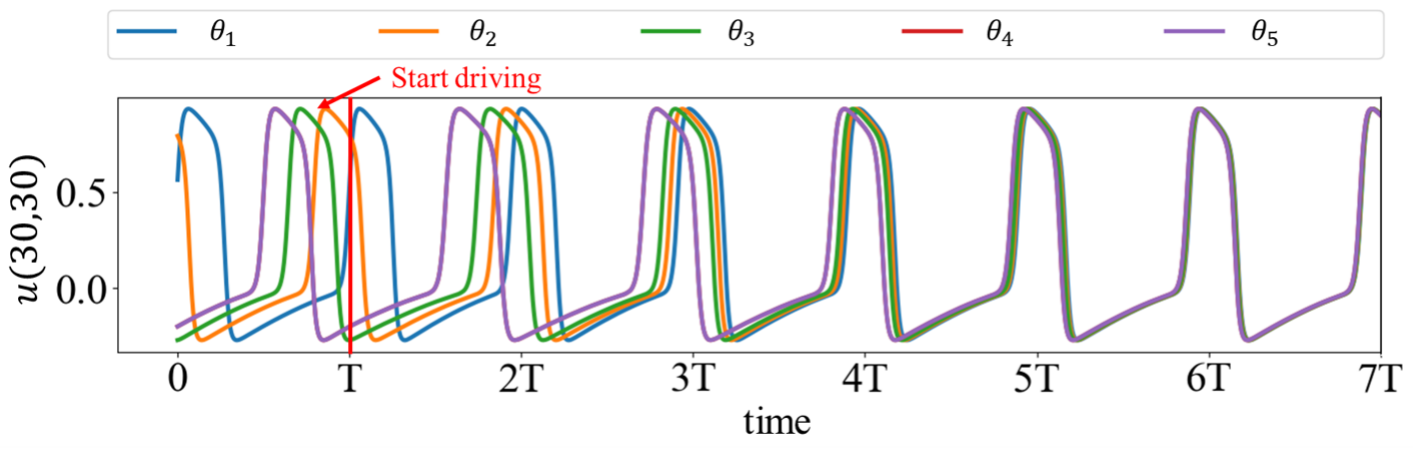}
\caption{Reference-based method for synchronizing five RD systems exhibiting spiral waves. Time sequences of the field variables $u$ at $(x, y) = (30, 30)$ of the five RD systems starting from different initial conditions on the limit cycle.}
\label{fig:spiral_ref}
\end{figure}

%
First, we present the results of synchronizing five RD systems with identical properties using the reference-based method.
The frequency $\omega_r$ of the reference phase is set to match that of the RD systems.
The driving input is applied from $t=T$, and the the driving intensity $\eta$ is set to $0.005$.
The initial phases of the spirals are given by $\theta_1 \simeq 0.18\pi$, $\theta_2 \simeq 0.58\pi$, $\theta_3 \simeq 0.88\pi$, $\theta_4 \simeq 1.17\pi$ and $\theta_5 \simeq 1.64\pi$ at $t=0$.
Figure~\ref{fig:spiral_ref} shows the evolution of the field variable $u$ at the position $(x, y) = (30, 30)$.
As in previous cases, synchronization of the five RD systems is successfully achieved.
The time necessary to achieve nearly complete synchronization is around $t \simeq 6T$.

\begin{figure}[htbp]
\includegraphics[width=\hsize]{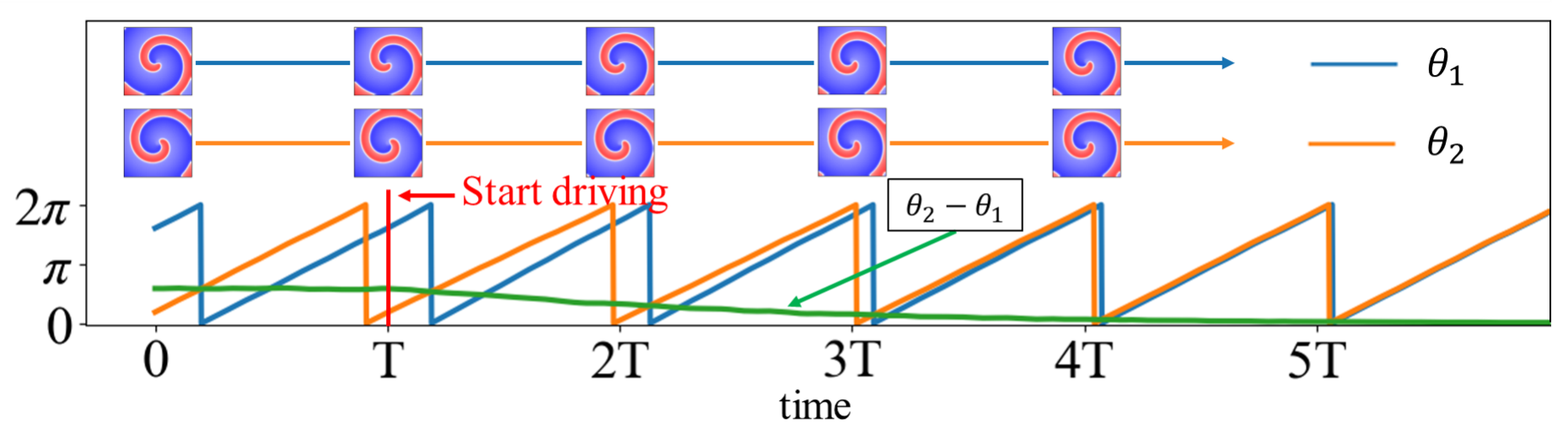}
\caption{Coupling-based method for synchronizing two RD systems exhibiting spiral waves. Evolution of the phases of the two RD systems (blue and orange), their phase difference (green), and the patterns of $u$ at $t = k T$ ($k=0, 1,2, ...$). 
}
\label{fig:spiral_coup}
\end{figure}

Finally, we present the results of synchronizing two RD systems using the coupling-based method.
The coupling input is applied from $t=T$, with the coupling intensity set to $\eta = 0.0025$.
The initial phases are set as $\theta_1 = 0$ and $\theta_2 \simeq 0.58\pi$ at $t=0$.
Figure~\ref{fig:spiral_coup} shows the evolution of the phases of the two systems, their phase difference, and the patterns of $u$.
We can confirm that the two RD systems achieve nearly complete synchronization within the duration of $4T$.

These results demonstrate that the phase autoencoder is also successful in synchronizing two-dimensional spiral wave patterns.

\section{Summary}

We have demonstrated that the phase autoencoder can extract reduced phase dynamics from high-dimensional rhythmic spatiotemporal patterns in a data-driven manner, without requiring prior knowledge of the underlying mathematical model. We proposed a data-driven phase control method
for RD systems using the phase autoencoder, which uses the decoder to compute the tangential direction of the limit cycle,  allowing the application of stronger driving inputs without inducing  amplitude deviations.

We applied the proposed phase autoencoder to two types of spatiotemporal patterns, i.e., the oscillating spot and spiral waves of the FitzHugh-Nagumo RD system, and confirmed that the phase autoencoder can successfully learn the asymptotic phase near the limit cycle and reconstruct the spatiotemporal patterns. We introduced two methods for synchronization of RD systems with tangential driving - the reference-based method and the coupling-based method. These methods do not require the evaluation of the phase and amplitude sensitivity functions, whose accurate estimation are challenging in RD systems due to high dimensionality. We also proposed a feedback control method to suppress the amplitude deviations of the system state from the limit cycle. We confirmed that the proposed methods can successfully synchronize the RD systems exhibiting limit-cycle oscillations.

In conclusion, our results demonstrate the feasibility and effectiveness of data-driven phase reduction and control of high-dimensional spatiotemporal dynamics through a low-dimensional latent-space representation obtained by the phase autoencoder.
As a final note, although we considered rhythmic RD systems as representative examples of spatiotemporally rhythmic systems in this study, the proposed phase-autoencoder and tangential driving method do not assume that the considered system is of the RD type. 
Therefore, they can be applied more generally to other rhythmic spatiotemporal systems, such as fluid flows.

\acknowledgments{K.Y., R.S. and H.N. acknowledge JSPS KAKENHI Nos. 25H01468,
25K03081, 22H00516, and 22K11919 for financial support. 
K.T. acknowledges the US AFOSR grant FA9550-21-1-0178 for financial support.
\bibliography{refs} 

\end{document}